# Dual-gated hBN/bilayer-graphene superlattices and the transitions between the insulating phases at the charge neutrality point

Takuya Iwasaki[1][†], Yoshifumi Morita[2], Kenji Watanabe[3], and Takashi Taniguchi[1]

[1]International Center for Materials Nanoarchitectonics, National Institute for Materials Science (NIMS), 1-1 Namiki, Tsukuba, Ibaraki 305-0044, Japan
[2]Faculty of Engineering, Gunma University, Kiryu, Gunma 376-8515, Japan
[3]Research Center for Functional Materials, NIMS, 1-1 Namiki, Tsukuba, Ibaraki 305-0044, Japan

[†]Corresponding author: IWASAKI.Takuya@nims.go.jp

Abstract
We report on transport properties in dual-gated hexagonal boron nitride (hBN)/bilayer-graphene (BLG) superlattices. Here, BLG is nontwisted, i.e., plain. This paper focuses on the charge neutrality point (CNP) for a plain BLG. Under a perpendicular magnetic field, transitions between two insulating phases at the CNP are detected by varying a displacement field with the study on the resistance-temperature characteristics and the magnetoresistance. This work opens avenues for exploring the global phase diagram of the hBN/BLG superlattices beyond the CNP.

In graphene and its bilayer counterparts, novel internal degrees of freedom ("flavor," e.g., valley) can play a crucial role when combined with a conventional "spin." In bilayer graphene (BLG), interaction-driven symmetry breaking is predicted due to its parabolic energy-band touching, where a finite density of states (DOS) is associated [1-6]. This contrasts with the vanishing DOS at nodal ("relativistic") energy-band touching in graphene. Such a finite DOS in BLG can trigger various types of instabilities. Previous studies focused on spontaneous gap opening at the charge neutrality point (CNP) of BLG [7-10]. In the quantum Hall regime at filling factor $\nu = 0$ of BLG, contrastingly, "magnetism" is generally expected whereby the system is spin and/or valley/layer polarized [7-19]. We also note more recent development of the quantum Hall phases in BLG beyond $\nu = 0$ [20, 21].

Furthermore, in BLG, the application of a perpendicular electric field allows for systematic control of the energy gap developing with a Mexican hat shaped band structure [22, 23], making it a promising candidate for future quantum devices [24, 25]. Such systematic engineering of the energy-band structure by a dual-gated device structure can lead to (hidden) ordered states. Here we want to add a "twist" to the above scenario by hexagonal boron nitride (hBN) [26, 27]. An ultraclean hBN/plain BLG device plays a crucial role in elucidating the fine structure of the phase diagram as a function of an electric and magnetic field. Due to the superlattice structure (moiré effects), even without an electric field, the alignment of the BLG with hBN harbors an energy gap, accompanied by nontrivial energy-band topology, i.e., accumulated Berry curvature (hot spot) [28], an energy band with a narrow bandwidth and van Hove singularities from the momentum space point of view. Furthermore, combined with the emergence of satellites of the CNP due to hBN, the Hofstadter butterfly takes place under a magnetic field [29, 30].

In this study, we investigated the transport properties in dual-gated hBN/BLG superlattices, focusing on the vicinity of the CNP. By mapping out the energy gap, we found that first-order phase transitions occur between two insulating states at the CNP under a perpendicular magnetic field. This is a demonstration of devices with a dual-gated structure and an hBN/BLG superlattice with a nearly zero-angle misalignment, and it will serve as a starting point for exploring the "global phase diagram" in the subsequent works, where the satellites; i.e., "the second-generation" quasiparticles play a crucial role.

Our experiment is conducted in the hBN/BLG(/hBN) device with top and bottom gates ("dual gate"), where hBN plays the role of high-quality dielectric on both sides. Figures 1(a) and 1(b) show our main device structures (Device I). BLG and hBN flakes were firstly prepared on a $SiO_2$/Si substrate by the mechanical exfoliation method from bulk crystals to fabricate the devices. We then assembled an hBN/BLG/hBN/graphite heterostructure by the bubble-free dry transfer method [31], which was etched into a Hall bar geometry by electron beam (EB) lithography and $CHF_3/O_2$ plasma. Electrodes with one-dimensional edge contact (Cr/Au) were fabricated by EB deposition [32]. Another hBN sheet was transferred onto the above device to protect the edge of the Hall bar. The top gate (Ti/Au) was then fabricated by EB lithography and deposition. The schematic cross-sectional view of our device is shown in Fig. 1(a). For the transport measurement, a four-terminal configuration with AC lock-in techniques (an excitation current $I \sim$ 10-100 nA and a frequency of 17 Hz) was applied to detect the longitudinal resistivity $\rho_{xx} = (V_{xx}/I)(W/L)$, and the Hall resistivity $\rho_{xy} = V_{xy}/I$, where $L$ and $W$ are the channel length and width, respectively. $V_{xx}$ and $V_{xy}$ are the voltages measured between the Hall bar electrodes shown in Fig. 1(b). The device was measured at low temperature ($T$) in a $^4$He cryostat with a superconducting magnet used for applying a perpendicular magnetic field ($B$). A graphite flake under the bottom hBN layer was used as the back gate. The top and back gates were used to modulate a vertical displacement field and the carrier density of the device independently. Because hBN can be grown as a crystal of high quality unlike an amorphous $SiO_2$, the charge traps are typically orders of magnitude suppressed compared to those in $SiO_2$. Thus, in the hBN/BLG structure, the influence of the impurities is reduced. Furthermore, in the graphite/hBN/BLG(/hBN) structure, the charged impurities in $SiO_2$ are further screened, and the charge fluctuations are suppressed drastically. Let us comment on the quality of our devices. The electron and hole Hall mobilities near the CNP are typically up to ~50,000 $cm^2$/V s and the mean free path is ~0.5 μm. Note that the CNP is detected around $V_{bg} \sim V_{tg} \sim 0$ V, which also implies the quality of our devices.

Figure 1(c) shows the $\rho_{xx}$ mapping as a function of top-gate ($V_{tg}$) and bottom-gate voltage ($V_{bg}$) at $T = 1.6$ K. $\rho_{xx}$ is modulated by dual-gating, by which we can independently tune the carrier density and the perpendicular displacement field. We define an average displacement field as $D = [C_{bg}(V_{bg} - V_{bg,0}) - C_{tg}(V_{tg} - V_{tg,0})]/2\varepsilon_0$, and the total carrier density as $n = [C_{bg}(V_{bg} - V_{bg,0}) + C_{tg}(V_{tg} - V_{tg,0})]/e$, where $C_{bg(tg)}$ is the back (top) gate capacitance per unit area, $V_{bg,0(tg,0)}$ is the offset of the back (top) gate voltage, $\varepsilon_0$ is the vacuum permittivity, and $e$ is the electron charge. The resistance peaks at the CNP and the satellites on the electron side (Sat-e) and the hole side (Sat-h) are shown in Fig. 1 (c), which demonstrates the formation of a superlattice structure due to the "twist" by hBN. From a distance between the CNP and the satellites (at $n \sim \pm 2.4 \times 10^{12}$ $cm^{-2}$), the moiré period is extracted to be ~ 13.8 nm, corresponding to the alignment angle of ~0.15° between BLG and hBN [30, 33]. It is known that two energy bands approaching, i.e., nearly touching, occurs with a finite energy gap at the CNP. This also applies to the satellites. This point is closely related to the "Hofstadter butterfly" under a magnetic field [29, 30].

As shown in Fig. 1(d), the Hall conductivity $\sigma_{xy} = -\rho_{xy}/(\rho_{xx}^2 + \rho_{xy}^2)$ for $B = 1-2$ T exhibits the plateaus at $\sigma_{xy} = 4N(e^2/h)$ (where $N$ is an integer, $h$ is the Planck's constant), demonstrating the feature of the quantum Hall effect of BLG [34]. Moreover, the "Hofstadter butterfly" is confirmed under a strong magnetic field, characteristic of the superlattice structure and the satellites (see Supplemental Material Fig. S1 [35]). We can observe the Landau fans not only at the CNP but also at the satellites, where peaks due to Landau levels are bundled near the zero magnetic field regime [29, 30]. This is a demonstration of devices with both dual-gated structures and hBN/BLG superlattice with a nearly zero-angle misalignment. Note that our devices are not "doubly" aligned, since we do not observe different superlattice periods larger than the maximum graphene/hBN moiré period [36].

Here we discuss the characteristics at $B = 0$. Figure 2(a) shows the $T$ dependence of the resistivity at the CNP ($\rho_{xx,CNP}$) for various $D$. The increase in $\rho_{xx,CNP}$ with decreasing $T$ indicates an insulating behavior, which is more pronounced at high $D$. In a certain temperature range (typically $T$ = 20-140 K, which depends on $D$; see Supplemental Material Fig. S6 [35]), the thermally activated behavior $1/\rho_{xx,\text{CNP}} \propto \exp(\Delta/2k_BT)$ is observed. The energy gap $\Delta$ is estimated from the Arrhenius plot by fitting $\ln(1/\rho_{xx,\text{CNP}}) \propto \Delta/2k_BT$ to the corresponding linear regime, as shown in the inset of Fig. 2(a). The resistance and energy gap evolve as a function of $D$, consistent in magnitude with previous results of BLG devices [22]. As shown in Fig. 2(b), the energy gap is tuned by the displacement field $D$. At the CNP/zero density, the map of the energy gap corresponds to the observation of a widely tunable band gap in BLG. Even without an electric field ($D$ = 0), alignment of the BLG with hBN harbors an energy gap $\Delta \sim 1.4$ meV in our Device I due to the superlattice structure (moiré effects). This is comparable with the previous result, ~1.7 meV [28]. It should be noted that the moiré effect (e.g., the energy gap) is a rapidly varying function of the misalignment angle between hBN and BLG, especially near zero angle [33]. Although the energy gap is tiny for a small $D$ regime and the thermally activated behavior has some ambiguity, the estimate in the small $D$ regime is consistent with an extrapolation from the large $D$ regime with a clear activation behavior.

The resistance at the satellites (in particular, Sat-h) can also show a thermally activated behavior with $\Delta \sim 6.0 \pm 0.2$ meV (Fig. 2(c)). Concerning Sat-e, only a weak (or no) signal for the gap is detected (Fig. 2(d)). This could be attributed to the high DoS at the Sat-e regime. In the graphene/hBN superlattice, the second-generation Dirac cones are singly degenerate for the valence band (Sat-h side), while they are triply degenerate for the conduction band (Sat-e side) [37]. This leads to the difference in the DOS between the Sat-e and Sat-h regimes. This is consistent with previous works on hBN/BLG devices and here we refer to one of the most recent approaches with a systematic control of the alignment angle between BLG and hBN [38]. The temperature dependence of the resistivity at both satellites is insensitive to $D$, in contrast to that at the CNP. Such an energy gap at the satellites shows characteristics due to the superlattice structure by hBN. We note that, since the thermally activation behavior is less clear here as commented on above, the details should be discussed in the future, combined with systematic theoretical works.

Next, we move on to the behavior of nonzero $B$. We shall apply a magnetic field there and map out the "phase diagram" at the CNP (see Supplemental Material, Fig. S2, for the mapping plot of $\rho_{xx}$ vs $V_{tg}$ and $V_{bg}$ near the CNP for nonzero $B$ [35]). As indicated by the phase diagram of the $\rho_{xx}$ mapping for $D$ and $B$ at the CNP and $T$ = 1.6 K (Fig. 3(a)), we assign the phase in both high $B$ and small $D$ regimes as the (quantum Hall) insulator with possible spin magnetism, which is not necessarily fully spin polarized, e.g., canted

(anti)ferromagnet [14-19]. This phase is enclosed by a white line in Fig. 3(a) with an enhanced conductance, with an insulating phase inside (red region).

We map out the energy gap over the phase diagram to discuss the transition. The energy gap, estimated by fitting to the Arrhenius plot with the same method as in the above discussion (Fig. 2 and Supplemental Material, Fig. S6 [35]), is also continuously decreased by a perturbation, i.e., applying $B$ as shown in Fig. 3(b). Contrastingly, there is a gap minimum at finite $B$ (e.g., Fig. 3(b) at $D = -40$ mV/nm), corresponding to a field-induced quantum phase transition between insulating states at charge neutrality. Figures 3(c)-3(e) show the $D$ dependence of $\rho_{xx}$ (lower panel) and $\Delta$ (upper panel) at fixed $B$. In lower $B$ (Figs. 3(d) and 3(e)), there is no signature of a continuously vanishing gap even in the vicinity of the criticality. Just on the criticality, in contrast to the insulating behavior in both sides/phases, a compressible "spike" takes place in the low-temperature limit reported in the literature [14-19]. This is consistent with a scenario of the formation of a microscopic network of the two competing phases at the first-order transition [39, 40]. Several studies observed an intermediate phase between the two insulating phases in higher magnetic fields [16, 18]. Also, as seen in Fig. 3(c) at $B = 7.5$ T, we detected a possible finite compressible region. This higher magnetic-field regime will be studied in the future. Note that disorder can obscure such a transition, and an ultraclean device is crucial to detect it.

Some comments are in order on the temperature dependence measurements of the phase around the field-induced phase transition with an increasing magnetic field. Under a high magnetic field, the energy gap defined above can be different from an "intrinsic" energy gap of the band structure. Furthermore, after the field-induced transitions, we do not necessarily observe a simple thermally activated behavior over a wide range of temperatures, although this phase is insulating. Instead, the energy gap is estimated temporarily by fitting the high-temperature regime as in ref. [14] (see Supplemental Material Fig. S6 [35]). It is also possible that the resistance-temperature ($R$-$T$) characteristics in this phase are irrelevant to the bulk energy gap. Under a finite magnetic field, this phase can have an edge mode whose gap is smaller than the bulk gap. If this is the case, the $R$-$T$ characteristics can be dominated by the edge gap, by which we define the energy gap in this phase. Moreover, in some of the literature, power-law behavior in the $R$-$T$ characteristic has also been reported [18], unlike in our case of ultraclean devices.

In summary, we have fabricated and characterized dual-gated devices based on hBN/BLG superlattices. Under a perpendicular magnetic field, transitions between two insulating states at the CNP are detected by mapping out the energy gap, deduced from the *R-T* characteristics and the magnetoresistance at the CNP. We have focused on the regime near the CNP in this paper. It would be interesting to extend this study beyond the CNP (discussed in the Supplemental Material, Sec. S3 [35], with comments on related devices [41, 42]) and access "higher-energy scales," e.g., near van Hove singularities, accompanied with a divergent DOS, and the satellites (the second-generation quasiparticle).

**Acknowledgments**
The authors thank H. Osato, E. Watanabe, and D. Tsuya from the NIMS Nanofabrication platform for assisting with the device fabrication. This work was partially supported by the Japan Society for the Promotion of Science (JSPS) KAKENHI Grant No. 21H01400, the NIMS Nanofabrication Platform in Nanotechnology Platform Program Grant No. JPMXP09F22NM5013, and the World Premier International Research Center Initiative on Materials Nanoarchitectonics sponsored by the Ministry of Education, Culture, Sports, Science and Technology (MEXT), Japan.

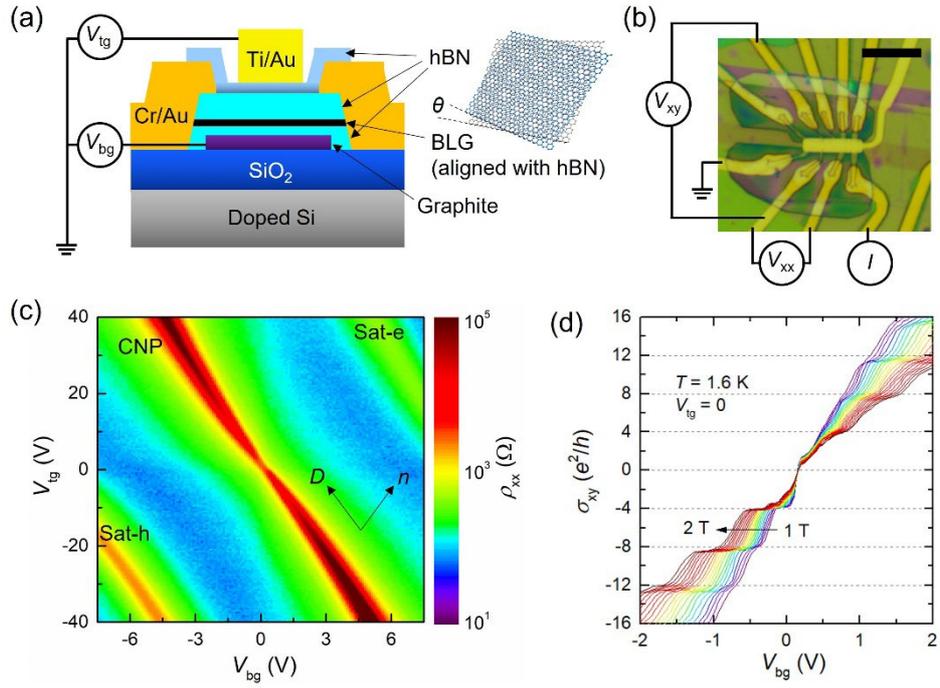

FIG. 1. (a) Schematic cross-section of the hBN/BLG superlattice device. The thicknesses of the passivation, top, and bottom hBN layers are 18, 52, and 46 nm, respectively, confirmed by atomic force microscopy. The crystalline axis of BLG is aligned with that of one of the nearest hBN layers. (b) Optical image of Device I, with the measurement setup. Scale bar, 10 μm. (c) Mapping plot of $\rho_{xx}$ vs $V_{tg}$ and $V_{bg}$ at $T$ = 1.6 K, $B$ = 0. (d) $\sigma_{xy}$ as a function of $V_{bg}$ for $B$ = 1-2 T at $V_{tg}$ = 0, $T$ = 1.6 K.

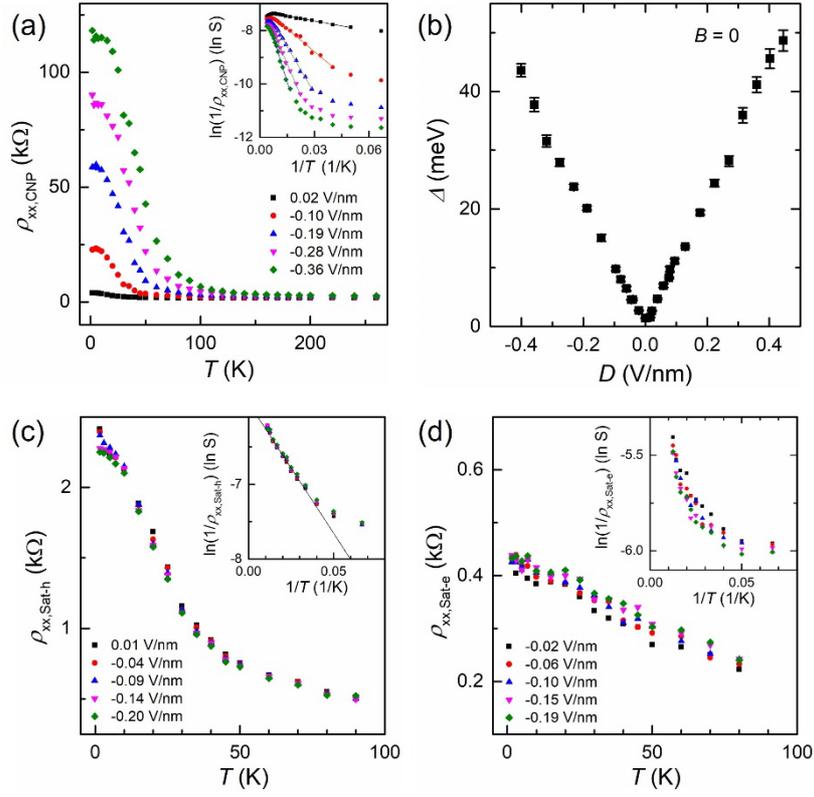

FIG. 2. (a) Temperature dependence of the resistivity at the CNP $\rho_{xx,\text{CNP}}$ for various $D$. Inset: the Arrhenius plot of $1/\rho_{xx,\text{CNP}}$. The symbols show the experimental data, while the solid lines correspond to the results of the fitting with $\ln(1/\rho_{xx,\text{CNP}}) \propto \Delta/2k_BT$. (b) Energy gap $\Delta$ as a function of $D$ at the CNP and $B = 0$. (c,d) $\rho_{xx}$-$T$ plot similar to (a) but for the resistivity at (c) the hole side $\rho_{xx,\text{Sat-h}}$ and (d) the electron side satellite point $\rho_{xx,\text{Sat-e}}$.

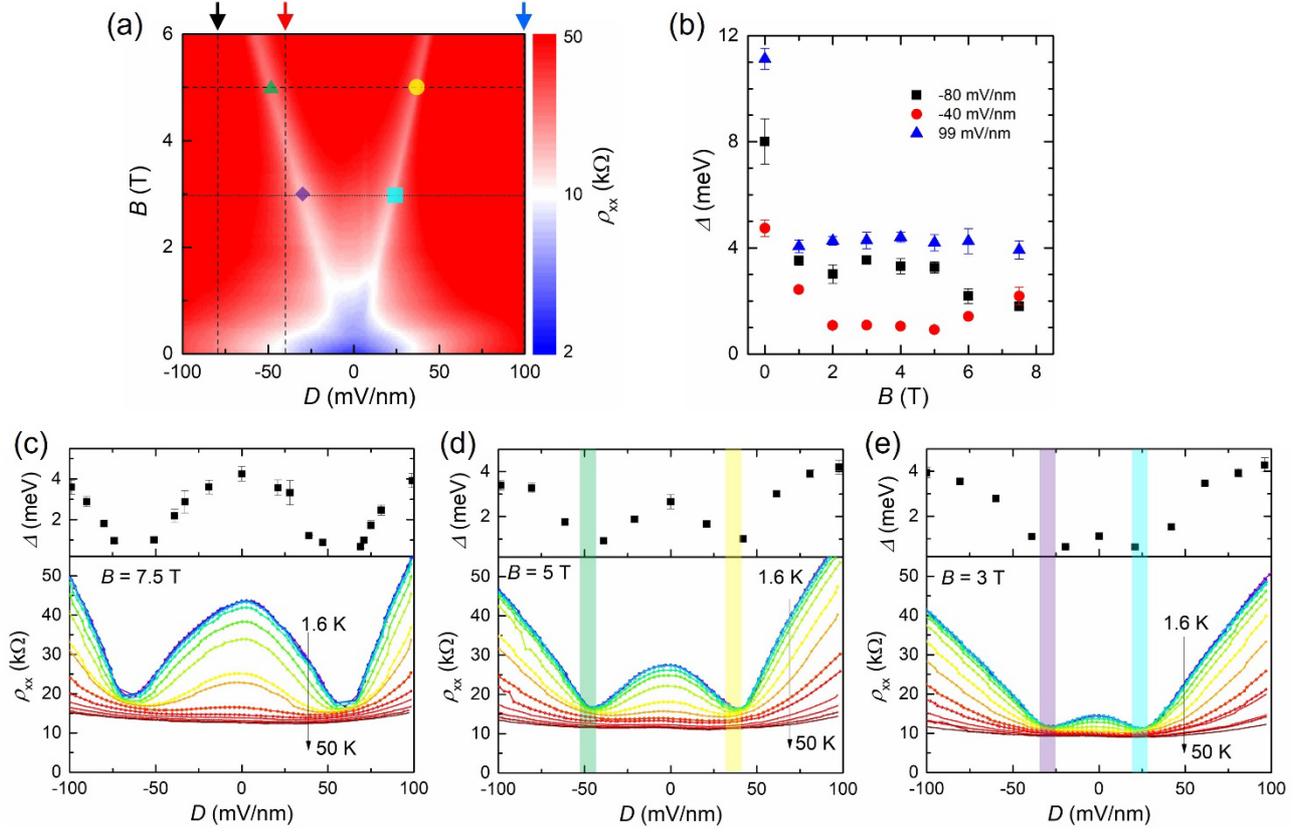

FIG. 3. (a) Mapping plot of the longitudinal resistivity $\rho_{xx}$ vs $D$ and $B$ at the CNP, $T = 1.6$ K. (b) Energy gap as a function of $B$ for $D = -80, -40,$ and $99$ mV/nm, corresponding to the vertical dashed lines indicated by the black, red, and blue arrows in (a), respectively. (c) Lower panel: $\rho_{xx}$ vs $D$ at $B = 7.5$ T. Upper panel: the energy gap $\Delta$ vs $D$ plot. The energy gap is estimated by fitting with the Arrhenius plot. (d,e) Same plots as (c) for (d) $B = 5$ T and (e) $B = 3$ T, corresponding to the horizontal dashed and dotted lines in (a), respectively. The colored stripes correspond to the marked points in (a).

# Supplemental Material for

**Dual-gated hBN/bilayer graphene superlattices and the transitions between the insulating phases at the charge neutrality point**

Takuya Iwasaki[1][†], Yoshifumi Morita[2], Kenji Watanabe[3], Takashi Taniguchi[1]

[1]International Center for Materials Nanoarchitectonics, National Institute for Materials Science (NIMS), 1-1 Namiki, Tsukuba, Ibaraki 305-0044, Japan
[2]Faculty of Engineering, Gunma University, Kiryu, Gunma 376-8515, Japan
[3]Research Center for Functional Materials, NIMS, 1-1 Namiki, Tsukuba, Ibaraki 305-0044, Japan

**Table of contents**

# S1. Butterflies and quantum Hall effect

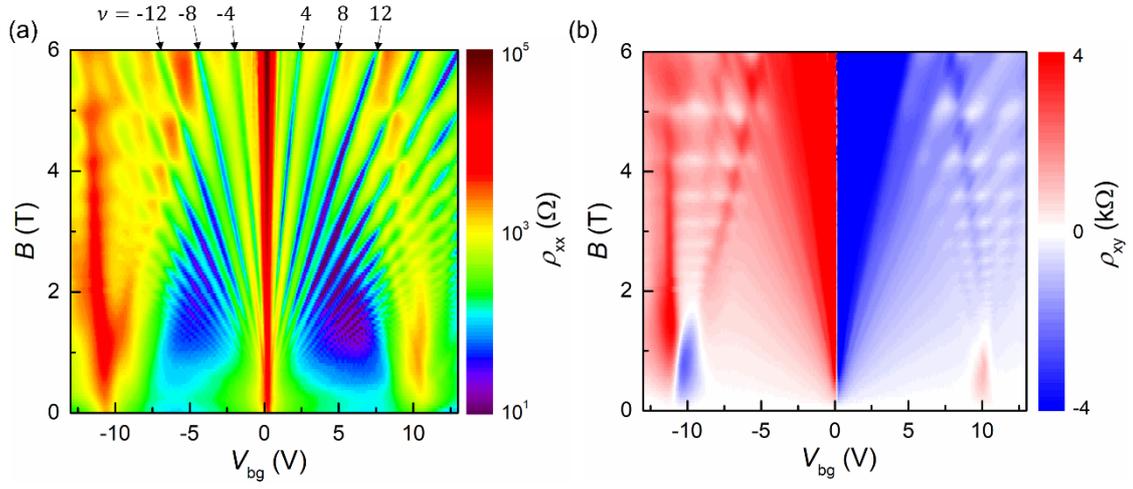

FIG. S1. Hofstadter's butterfly spectra. (a) Mapping plot of the longitudinal resistivity $\rho_{xx}$ vs. back-gate voltage $V_{bg}$ and a perpendicular magnetic field $B$ at the temperature $T = 1.6$ K and top-gate voltage $V_{tg} = 0$. The arrows indicate the lines where the filling factor $v$ is constant. (b) Same plot as (a) for the transverse resistivity $\rho_{xy}$.

## S2. More on $\nu = 0$

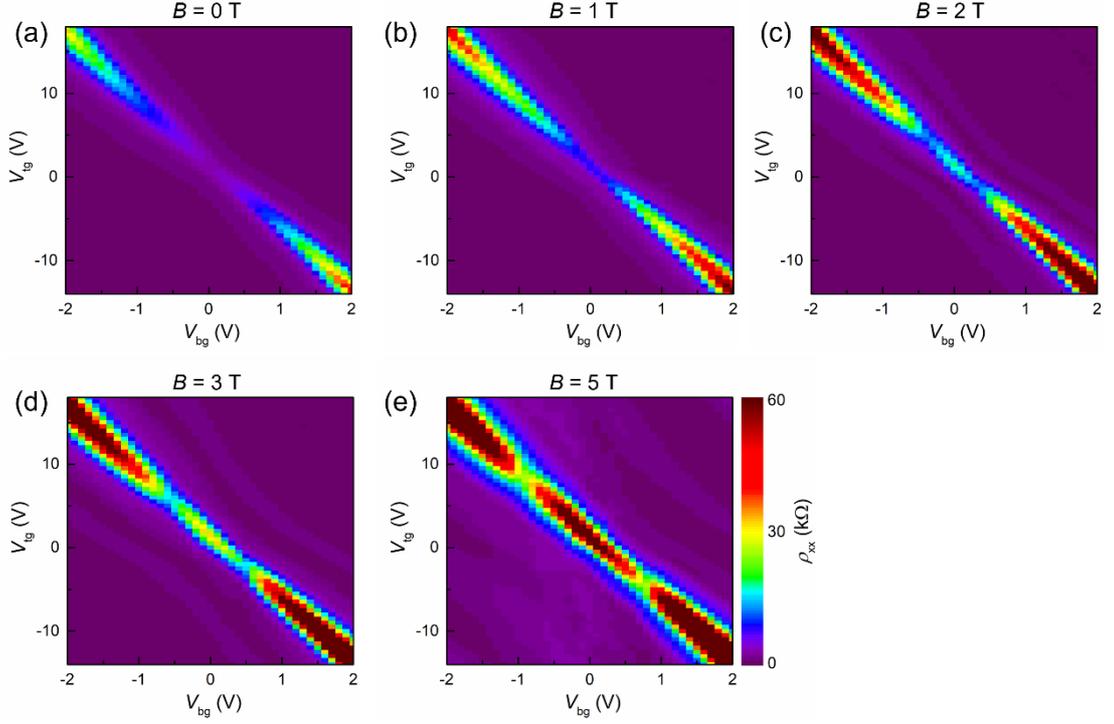

FIG. S2. Mapping plot of $\rho_{xx}$ vs. $V_{bg}$ and $V_{tg}$ at $T = 1.6$ K. The gate voltage range is focused on the charge neutrality point (CNP) in a small displacement field. (a) $B = 0$, (b) 1 T, (c) 2 T, (d) 3 T, and (e) 5 T. As $B$ increases, the high resistive region corresponding to the spin-polarized insulating phase is gradually developed near the center of the plot (small displacement field $D$).

## S3. Away from the CNP /Another device (a reproduction)

Here, let us show some data beyond the focus of the main text to be complementary to it. For example, in Fig. 2 of the main text, we focused only on the resistance-temperature ($R$-$T$) characteristics at the CNP/satellites. Moreover, in Fig. 3 of the main text, we focused on the magnetoresistance only at the CNP. Here we show the $R$-$T$ and the magnetoresistance characteristics away from the CNP in Fig. S3 and Fig. S5, respectively. A reproduction on another device (Device II) is also included (Fig. S4). We focus on two dual-gated hBN/BLG superlattices (Device I and II). As a reference, we also comment on more hBN/BLG superlattices (see, for example, [S1, S2]). Note that the devices live on multi/higher dimensional "parameter space," which is projected onto our results, and several key ingredients are "entangled."

Away from the CNP, in the high-temperature regime, conventional phonon contribution is observed in the $R$-$T$ in all devices, as shown in Fig. S3 and S4 (plus "Umklapp effects" due to the Coulomb interaction at some region in the phase diagram) [S1]. In a lower temperature regime, the $R$-$T$ in some devices shows a departure from such conventional behavior (Fig. S3 and S4), i.e., resistance "bump" (and "drop" nearby) in the terminology of ref. [S2]. Whether such "bump" (and/or "drop") survives depends on the detailed conditions commented below. This can be due to some "signatures," e.g., critical fluctuations of "hidden order."

Initially, we focus on the alignment angle between BLG and one of the two hBN layers. In the range from ~0° to ~0.2° in the ultra-clean devices, we have got qualitatively similar outputs (sometimes "bump" is absent, as shown in Fig. S3). Contrastingly, in the case of a device (of moderate quality) with an alignment angle of ~0.9°, we could not get such a behavior [S1]. Note that there is a spatial fluctuation in this angle and physical properties, e.g., an energy gap is a rapidly varying function of this angle, especially near zero angle. One more comment is in order. BLG is sandwiched between two hBN layers, i.e., hBN/BLG/hBN superlattices in our devices. Here the global character of, e.g., the Hofstadter butterfly is the same as in ref. [S3], where one of the two hBN layers is aligned to the BLG. However, the other hBN is largely misaligned (~15°) in ref. [S3]. Contrastingly, in our devices, the other hBN is not so much misaligned/nearly aligned in our devices, at least on the edge orientation. Though, note that our devices are not "doubly" aligned in the sense of the terminology in ref. [S4], since we do not observe features of doubly well-aligned devices, such as different superlattice periods larger than the maximum graphene/hBN moiré period [S4]. In some devices, though not necessarily central, this point can play some role in controlling the "detail," e.g., electron-phonon coupling, etc. As shown in Fig. S3 and S4, the slope of $R$-$T$ characteristics in the high-temperature regime can depend on such a detail, which should be basically due to conventional phonon contribution in a narrow energy band with van Hove singularities [S1].

Next, let us show the magnetoresistance at $T = 1.6$ K away from the CNP. For example, near the van Hove singularity, an extremely large magnetoresistance is revealed with a V-shaped form, as shown in Fig. S5 [S1]. This will be discussed in detail in a different paper with the $R$-$T$ characteristics.

Figure 1 of the main text shows our main device structures, hBN/BLG/hBN heterostructures with a Hall bar geometry. Another device (Device II, shown in Fig. S4(a)) is also measured and plays a complementary role, which reconfirms our main results (Fig. S4(b)). Device II has a moire period of ~ 10.0 nm and the angle alignment between BLG and hBN of ~0.99°. Here in this Supplemental Material, we have also shown some outputs of the Device II, i.e., $R$-$T$ characteristics and magnetoresistance at the CNP and away from there (Fig. S4). In particular, the magnetoresistance at the CNP (Fig. S4(b)) is a reproduction of the results in the main text (note that the quantitative feature should be different between device I and II since the energy gap is a rapidly varying function of the misalignment angle between hBN and BLG).

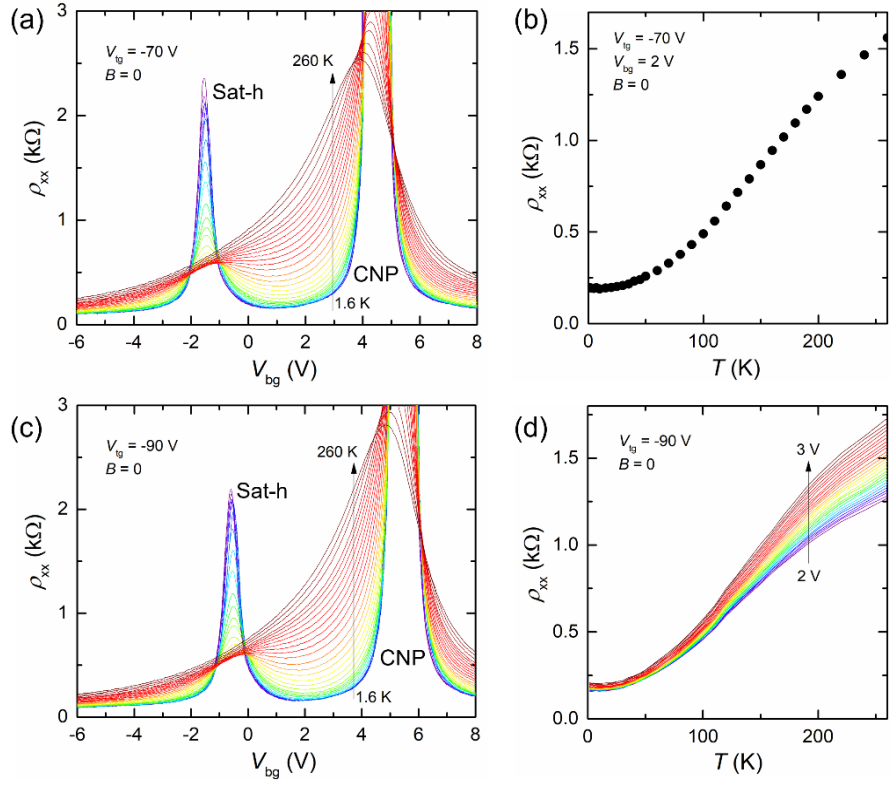

FIG. S3. (a) $\rho_{xx}$ vs. $V_{bg}$ plot at $V_{tg} = -70$ V for various $T$. (b) $\rho_{xx}$-$T$ characteristic at $V_{tg} = -70$ V, $V_{bg} = 2$ V. (c) Similar plot as (a) but at $V_{tg} = -90$ V (higher $D$). (d) Similar plot as (b) but at $V_{tg} = -90$ V, for $V_{bg} = 2$-$3$ V.

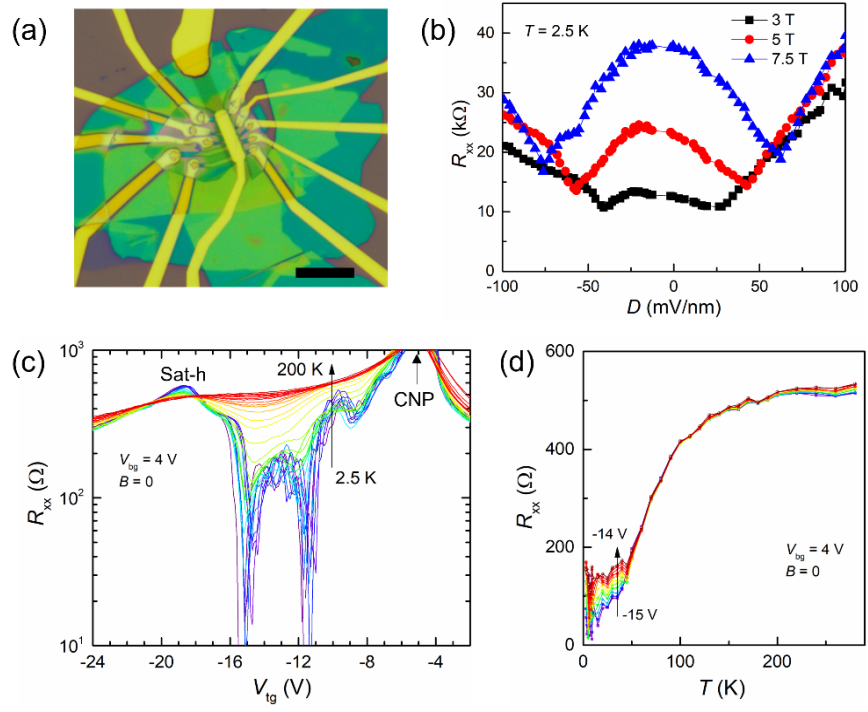

FIG. S4. Another device: Device II. (a) Optical image of the device. Scale bar, 10 μm. (b) Longitudinal resistance $R_{xx}$ vs. $D$ for $B = 3, 5$, and 7.5 T at $T = 2.5$ K. (c) $R_{xx}$ as a function of top-gate voltage $V_{tg}$ at $V_{bg} = 4$ V for various $T$. (d) $R_{xx}$-$T$ plot at $V_{bg} = 4$ V for $V_{tg}$ from –15 to –14 V.

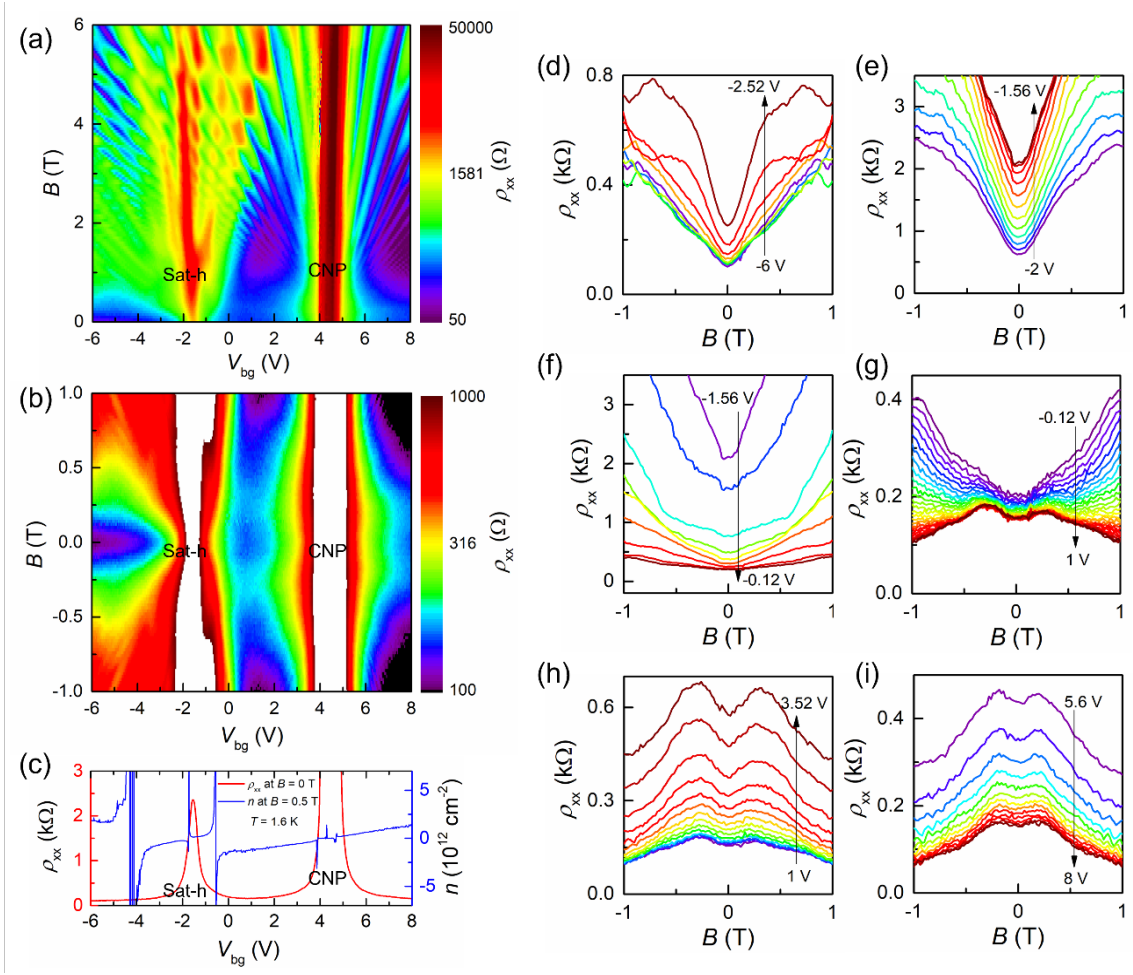

FIG. S5. Magnetotransport. (a,b) Mapping plot of $\rho_{xx}$ vs. $V_{bg}$ and $B$ at $T = 1.6$ K and $V_{tg} = -70$ V for (a) high $B$ and (b) small $B$ range. (c) $\rho_{xx}$ and the carrier density $n$ as a function of $V_{bg}$ at $T = 1.6$ K and $V_{tg} = -70$ V. $\rho_{xx}$ and $n$ are measured at $B = 0$ T and 0.5 T, respectively. (d-i) Line-cut plots for (b) at each $V_{bg}$ range.

## S4. Fitting to Arrhenius: an example

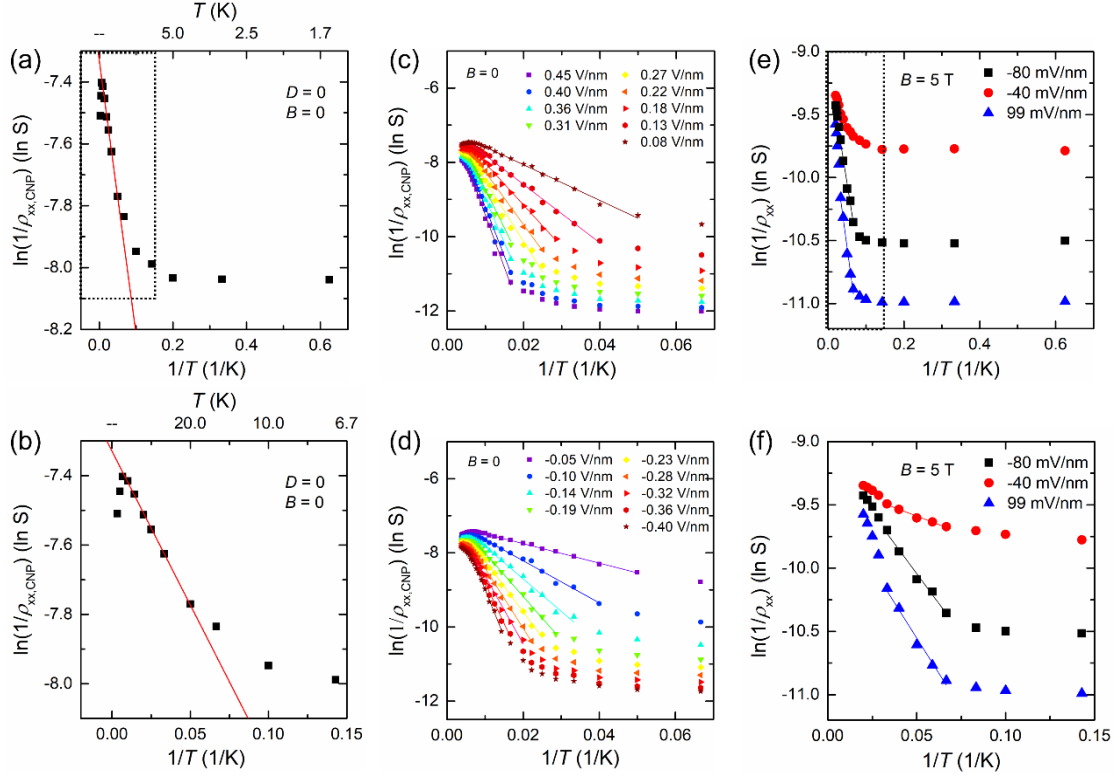

FIG. S6. Arrhenius plots, $\ln(1/\rho_{xx})$ vs. $1/T$. The symbols show the experimental data, while the solid lines correspond to the results of the fitting with $\ln(1/\rho_{xx,\text{CNP}}) \propto \Delta/2k_BT$. (a) At $B = D = 0$. The fitting range is $T = 20\text{-}150$ K. (b) Zoom up of the dotted box in (a). (c,d) At $B = 0$ and (c) $D > 0$, (d) $D < 0$. The fitting range is $T = 20\text{-}140$ K. (e) At $B = 5$ T and $D \neq 0$. The fitting range is $T = 15\text{-}30$ K. (f) Zoom up of the dotted box in (e).